# Thermal Conductivity of Suspended Few-Layer MoS$_2$

Adili Aiyiti[1,2,3]*, Shiqian Hu[1,2,3]*, Chengru Wang[4]*, Qing Xi[1,2,3], Zhaofang Cheng[5,6], Minggang Xia[5,6], Yanling Ma[4], Jianbo Wu[4], Jie Guo[1,2,3], Qilang Wang[1,2,3], Jun Zhou[1,2,3], Jie Chen[1,2,3], Xiangfan Xu[1,2,3], Baowen Li[7]

[1]Center for Phononics and Thermal Energy Science, School of Physics Science and Engineering, Tongji University, 200092 Shanghai, China

[2] China-EU Joint Center for Nanophononics, School of Physics Science and Engineering, Tongji University, 200092 Shanghai, China

[3]Shanghai Key Laboratory of Special Artificial Microstructure Materials and Technology, School of Physics Science and Engineering, Tongji University, Shanghai 200092, China

[4] School of Materials Science and Engineering, Shanghai Jiaotong University, 200240 Shanghai, China

[5]Department of Optical Information Science and Technology, School of Science, Xi'an Jiaotong University, 710049 Xi'an, China

[6] Department of Applied Physics, School of Science, Xi'an Jiaotong University, 710049 Xi'an, China

[7]Department of Mechanical Engineering, University of Colorado, Boulder, CO 80309-0427, USA

Modifying phonon thermal conductivity in nanomaterials is important not only for fundamental research but also for practical applications. However, the experiments on tailoring the thermal conductivity in nanoscale, especially in two-dimensional materials, are rare due to technical challenges. In this work, we demonstrate *in-situ* thermal conduction measurement of MoS$_2$ and find that its thermal conductivity can be continuously tuned to a required value from crystalline to amorphous limits. The reduction of thermal conductivity is understood from phonon-defects scatterings that decrease the phonon transmission coefficient. Beyond a threshold, a sharp drop in thermal conductivity is observed, which is believed to be a crystalline-amorphous transition. Our method and results provide guidance for potential applications in thermoelectrics, photoelectronics, and energy harvesting where thermal management is critical with further integration and miniaturization.

* These authors contributed equally to this work.

Correspondence and requests for materials should be addressed to X.X. (email: xuxiangfan@tongji.edu.cn ) or to B.L. (email: Baowen.Li@colorado.edu ).

# Introduction

Manipulating and tailoring the electrical and thermal properties are required in nano electronic and thermoelectric applications.[1] Theoretical works have predicted the role of defects in modulating the electrical transports by generating defect-induced localized states.[1-3] However, different from controlling electrons or photons, controlling phonons or heat is much more challenging, especially in the nano-scale. In fact, tailoring or even measuring the thermal conductivity of suspended and extremely fragile thin film such as two-dimensional materials (2D) is even more difficult. Only few experimental works on modulation of thermal conductivity of graphene by phonon-boundary scattering,[4] high energy beam irradiation[5,6] and phononic crystal[7] are reported so far. However, these pioneering works are demonstrated in different samples, which ignore sample-to-sample variations and thermal contact resistance.

Different from graphene, the layered transition metal dichalcogenide (TMDC), $MoS_2$, is proposed as a promising potential candidate for photoelectrics and thermoeletrics due to its layer-dependent gap.[8-10] A material that behaves as an electron-crystal and phonon-glass is an ideal thermoelectric material, which allows a relatively large temperature gradient across it while conducting electrons efficiently to generate a thermoelectric voltage.[11] Namely, a higher Seebeck coefficient and a lower thermal conductivity are preferred to achieve a considerable figure of merit. As a consequence, defect engineering[12,13] on $MoS_2$ provides a new approach to modulate their properties[14] or even create novel functionalities.[1,15] Electrical[16] and optical[17] properties of a single layer $MoS_2$ as well as the electronic transport properties[18] of few layer $MoS_2$ were tailored by controlled plasma. However, to the best of our knowledge, the experimental work on engineering the thermal conductivity of $MoS_2$ is still missing, although some theoretical and experimental study of thermal conductivity of $MoS_2$ have been reported.[19-23]

In this paper, we experimentally measured the thermal conductivity of suspended (exfoliated) few-layer $MoS_2$ transferred by improved dry transfer method,[24] by which organic contamination is avoided. More importantly, we demonstrated a novel approach to continuously tailor the thermal conductivity bit by bit by applying mild oxygen plasma. By controlling the oxygen plasma exposure time, we managed to tune the thermal conductivity of $MoS_2$ to a desired value. The thermal conductivity reduction under diluted defects is understood from the phonon-defects scatterings induced decrease of phonon transmission coefficient. A crystalline-amorphous transition emerges under high-dose plasma with a sharp change in the thermal conductivity. The temperature dependent thermal transport measurements combining the minimum conductance calculation indicate that thermal conductivity in heavily doped MoS$_2$ are approaching the amorphous limit.

**Results and Discussions**

Exfoliated few-layer MoS$_2$ flakes were carefully transferred onto the pre-patterned suspended thermal bridge devices by dry transfer method. Compared to the samples that were fabricated by PMMA mediated wet-transfer method (Supplementary Information Part 1), the samples prepared by dry transfer method are believed to have superior quality due to less polymer residues on surfaces.[24] As prepared samples were annealed to clean the possible residues on the surfaces and to enhance the contact before any thermal measurement. The thickness and vibrational status of the samples were confirmed by Raman Spectroscopy (Fig.1(a)),[25] while the length and width as well as the surface and edge status of the samples were characterized by SEM (inset of Fig.1(b)). Photoluminescence Spectroscopy (PL) data is given in Supplementary Information Part 1. Parameters of samples used for thermal measurement are given in table 1. We employed the pre-patterned suspended thermal bridge method and similar process for thermal measurements.[26-29] (Measurement details are described in the Supplementary Information Part 2.).

Fig. 1(b) shows the measured thermal conductance of the three samples that are suspended on the thermal bridge. The two samples, MoS$_2$-B and MoS$_2$-C, which have similar dimension show similar thermal conductance. The third one, MoS$_2$-A, which is wider and thicker than MoS$_2$-B and MoS$_2$-C, shows relatively larger thermal conductance values. The measured thermal resistance ($R_s$) of the MoS$_2$ samples contains the diffusive thermal resistance of the suspended section ($R_d$) and the thermal contact resistance between MoS$_2$ and contacting electrodes ($R_c$), which are given by following equations:

$$R_s = R_d + R_c \tag{1}$$

$$R_d = \frac{L}{\kappa t W} \tag{2}$$

$$R_s W = \frac{L}{\kappa t} + R_c W \tag{3}$$

where $\kappa$, $L$, $t$, and $W$ are the thermal conductivity, length, thickness, and width of the suspended MoS$_2$, respectively. The $R_d$ value decreases with increasing $t$ and decreasing $L$ so that $R_c$ can be derived by taking the limit of $L/t \rightarrow 0$. As shown in Fig. 1(c), $R_sW$ values of all three samples are plotted as a function of $L/t$, which is based on the assumption that the same thermal contact resistance per unit contact area for all samples. $R_c$ values are derived from the $R_cW$ that correspond to the intercept of the linear fitting at $L=0$. The strictly linear behavior in Fig.1(c) shows a uniformly sample quality between these three measured MoS$_2$ devices and further indicate the reliability of dry transfer method in preparing samples suitable for thermal measurements. The obtained $R_c$ and $R_s$ values are plotted as a function of temperature in Fig. 1(d). The obtained room-temperature $R_c$ values are 48%, 26% and 19% of the measured $R_s$ for the three samples whose suspending lengths are 1$\mu m$, 2$\mu m$ and 3$\mu m$ respectively.

Moreover, the electron-beam self-heating technique[30] is applied to double check the thermal contact resistance and intrinsic thermal conductivity. The derived thermal conductivity for the MoS$_2$-A, MoS$_2$-B and MoS$_2$-C are 30±3 Wm$^{-1}$K$^{-1}$, 34±5 Wm$^{-1}$K$^{-1}$ and 31±4 Wm$^{-1}$K$^{-1}$, respectively. The measured room-temperature thermal conductivity of few-layer MoS$_2$ is similar with some results in previous works.[19, 31, 32] By contrast, a higher value (44 Wm$^{-1}$K$^{-1}$) measured by similar suspended thermal bridge method for 4 layer MoS$_2$ was reported by Insun Jo et al.[20] earlier. Besides, an even larger value (52 Wm$^{-1}$K$^{-1}$) was reported by S. Sahoo et al.[22] This may result from the different sample preparing process, sample dimensions and measurement method. Compared with the thick MoS$_2$ deposited on the substrate by a modified high-temperature vapor-phase method,[33] our samples are exfoliated from the bulk MoS$_2$ and transferred to the thermal bridge devices by dry transfer method, during which some defects, rough edges are induced. Both defects and rough edges decrease the intrinsic thermal conductivity of MoS$_2$ on some degree. On the other hand, different MoS$_2$ crystal seeds containing different grain size may initially limit the intrinsic thermal conductivity. One interesting fact is that all the experimental values including our results for the thermal conductivity of few-layer MoS$_2$ are smaller than the ones (~100 Wm$^{-1}$K$^{-1}$) for its bulk counterpart reported by Liu et al.[34] This may related to the fact that the in-plane thermal conductivity of multilayer MoS$_2$ is insensitive to the number of layers,[35] for the finite energy gap in the phonon spectrum of MoS$_2$ makes the phonon–phonon scattering channel almost unchanged with increasing layer number. Instead, the defects, rough surface and edges in the few-layer ribbons emphasized above probably lowers the thermal conductivity. Even, with the existing experimental data combined together, it is still insufficient to conclude the trend of layer dependency of thermal conductivity for $MoS_2$. It is important to note that thermal conductivity in other 2D materials, e.g. graphene, boron nitride and black phosphorous etc., are not settled due to reasons mentioned above.[36]

In order to tailor the thermal conductivity of few-layer $MoS_2$, we take advantage of the mild oxygen plasma (FEI Plasma Cleaner Unit). The *in-situ* thermal conductivity measurement was carried out inside SEM chamber with build-in measurement stage and oxygen plasma cleaner. Fig. 2(a) shows the mimic diagram of the mild oxygen plasma treatment. With the assumption that same exposure time produces same amount of plasma and the plasma interacts homogeneously with per unit area, the exposure time was taken as the horizontal axis of the plot directly for the simplicity and clarity. Thus, the thermal conductivity versus exposure time curves is plotted in Fig. 2(b). The inset shows zoomed-in view of the Fig. 2(b) in the time range of 45~80 min. The thermal conductivity vs. plasma exposure time curves for both $MoS_2$-B and $MoS_2$-C follow similar regulations. The whole curve is separated into three regions by the two characteristic turning points.

**a) $t \leqslant 2$ minutes**: After 1 minute only, namely after very limited point defects are induced by the oxygen plasma, the thermal conductivity decreases rapidly. As clearly shown in Fig. 2(c), a quite obvious normalized thermal conductivity reduction rate is found for the first two minutes. The reduction of room-temperature thermal conductivity by small number of defects is about 14% and 36% for $MoS_2$-B and $MoS_2$-C, respectively. Some tiny ripples might be induced during the transferring. These ripples are likely to be covered by amorphous carbon induced during the sample characterization in the SEM chamber. When exposed to the oxygen plasma, the surface of the $MoS_2$-B is protected for a while by the coated amorphous carbon somehow. Thus, it seems reasonable that $MoS_2$-B encounters relatively lower normalized thermal conductivity reduction rate of 14% and 9% for the first two minutes respectively.

**b) $2 < t \leq 40$ minutes:** In this region two competing mechanisms dominate the thermal conductivity. After certain exposure time, corresponding amount of Sulphur atoms are dislodged and the induced vacancies result in thermal conductivity reduction due to the enhanced phonon-defect scatterings. Meanwhile, the partial vacancies and dangling chemical bonds are healed by the oxygen ions in some level.[16, 37] But, this kind of healing seems not robust and stable enough to resist the secondary damage effectively. Thus, the thermal conductivity decreases slowly with the increasing exposure time.

**c) $t > 40$ minute:** After quite long time, the thermal conductivity approached an asymptotic value. It is worth to note that a sharp jump in thermal conductivity is clearly shown in Fig. 2(b) and Fig. 2(d) (The normalization of exposure time into defect concentration is discussed in Supplementary Information Part 3.), which indicates a possible phase transition in $MoS_2$. We believe this phase change is related to the crystalline-amorphous transition of the $MoS_2$ resulting from the mild oxygen plasma. The thermal conductivity approached slowly to an asymptotic value. According to the minimum thermal conductivity model, the phonon mean free path cannot become arbitrarily short as the scattering strength increases.[38] With the increasing of the time of plasma treatment, the increased phonon-defect scattering explains the decreasing of thermal conductivity at the beginning. Once the phonon mean free path reaches its lower limit, the phonon-defect scattering is at its maximum effectiveness and no further reduction in thermal conductivity is possible. As a result, the thermal conductivity is saturated when the normalized defect concentration is above 80%.

To understand the underlying physical mechanism of the decrease of thermal conductivity by applying the mild plasma in experiment, we investigate the defects effect on the thermal transport in pristine four-layers $MoS_2$ to mimic the experimental system via non-equilibrium molecular dynamics (NEMD) simulations (Fig. 3(a)). As shown in Fig. 3(b), the thermal conductivity of the defected $MoS_2$ ($\kappa_{DMoS2}$) are significantly reduced compared to the thermal conductivity of $MoS_2$ ($\kappa_{MoS2}$), and decreases monotonically with the defect concentration increasing from 0.5% to 5%. In addition, the thermal conductivity is obviously more sensitive to the defects when the defects concentration is low. For instance, $\kappa_{DMoS2}$ is reduced to 55% of $\kappa_{MoS2}$ when the defect concentration is 0.5%. However, after further increasing the defect concentration from 0.5% to 5%, the $\kappa_{DMoS2}$ only decreases another 32%. This behavior is consistent with the previous MD simulations[39, 40] and agrees with our experimental results (Fig. 2(d)). Interestingly, our MD results (Inset for Fig. 3b) show that thermal conductivity of both pristine and defective $MoS_2$ is insensitive to the number of layers. The thermal conductivity didn't show obvious layer dependency.

We estimate the spectral phonon transmission coefficient of pristine $MoS_2$ and the defected $MoS_2$ with different defect concentration following the Ref.[41, 42]. As shown in Fig. 3(c), we find that when the defects are introduced, except for the zone-center extremely low-frequency phonons, the phonon transmission coefficient of the defected $MoS_2$ (dashed and dashed dot line) are decreased significantly compared with that of pristine $MoS_2$ (solid line) for the nearly entire frequency range. A further decrease is observed in the defected MoS$_2$ as the defect concentration increases from 0.5% to 5%. Based on our simulations, the phonon-defect scatterings should be responsible for the decrease of phonon transmission coefficient and the reduction of the thermal conductivity.

The exotic crystalline-amorphous transition is further evidenced by HRTEM characterization as well as the typical characteristic of temperature-dependent thermal conductivity of crystalline and amorphous phase of the samples. Fig. 4(a) and Fig. 4(b) show the HRTEM images and diffraction patterns of the intrinsic sample and tailored (after plasma treatment) sample, indicative of crystalline and amorphous phase respectively. The obtained thermal conductivity (before any plasma treatment) of the samples is plotted as a function of temperature in logarithmic scale in Fig. 4(c). With the increasing of temperature, the thermal conductivity of MoS$_2$ initially increases due to the activation of more phonon modes and reaches its peak value around $T$=125 K before the thermal conductivity decreases due to the increased Umklapp scattering with further increasing temperature, which indicates that the pristine samples are in typical crystalline phase. (The thermal conductivity of MoS$_2$ prepared by wet transfer method is also plotted for comparison.)

In contrast, the thermal conductivity of MoS$_2$-B after whole oxygen plasma treatment are shown in Fig. 4(d). The inset for (d) shows the measured thermal conductance of MoS$_2$-B, which is 1~2 order of magnitude larger than the background heat conductance of the blank suspended device at room temperature. Thermal conductivity increases monotonically with temperature between 20 K and 300 K with no observable peak, in sharp contrast with that in pristine samples. Furthermore, the measured thermal conductivity in MoS$_2$-B is two orders of magnitude smaller, e.g. thermal conductivity in MoS$_2$-B is around 1 Wm$^{-1}$K$^{-1}$ at $T$=300 K, which is even smaller than that in amorphous SiO$_2$. The extinct peak at low temperature together with tremendous reduction in thermal conductivity imply that the samples are in amorphous phase and verifies that the steep jump mentioned above originates from the crystalline-amorphous transition of the MoS$_2$.

To give further evidence that the samples are in the amorphous state after highly oxygen plasma dose, we adopt the modified anisotropic minimum thermal conductivity model proposed by Chen and Dames[43] to estimate the theoretical low limit of amorphous thermal conductivity in MoS$_2$. The calculated in-plane minimum thermal conductivity is plotted in Fig. 4(d) by dashed line. The theoretical line agrees well with the experimental data of MoS$_2$-B when $T>200$ K.

## Conclusions

We experimentally investigate the thermal conductivity of the few-layer MoS$_2$ by suspended thermal bridge method. By applying mild oxygen plasma, we also demonstrate a novel and effective method to tune the thermal conductivity of the MoS$_2$ to a desired value between its intrinsic value and amorphous limit, during which a crystalline-amorphous transition is observed. Our investigation here provides physical insights of engineering thermal property of MoS$_2$ and may shed lights in the applications of MoS$_2$ in thermal management and control.

## Experimental Section

*Device fabrication:* The suspended devices suitable for thermal bridge method measurement were obtained by standard nanofabricating process similar with that in previous works.[26, 28] A 500nm-thick low-strain SiN$_x$ film was deposited on the silicon wafer by chemical vapor deposition (CVD) method, followed by a standard deep UV photolithography, metal deposition and lift-off process. The reactive ion etching RIE was applied to etch exposed SiN$_x$ which was not covered by the pre-patterned Pt electrodes. At last, the exposed silicon was etched away by wet etching method, to release the suspended structures.

*Sample characterization:* The length and width were measured with FEI Nova Nano SEM 450. The number of layers was determined by HR800 Raman Spectroscopy, according to the layer-dependent Raman shift of MoS$_2$. The samples (before and after plasma treatment) were characterized by taking transmission electron microscopy (TEM) images, high-resolution TEM (HRTEM) images and selected area electron diffraction (SAED) using a JEOL JEM-2100F microscope with an accelerating voltage 200 KV.

*Thermal conductivity measurement:* Details of suspended thermal bridge method are discussed in Supplementary Information.

*Molecular dynamics simulations:* By referring to the Ref.[44] and Ref.[45] MD simulations in this paper are performed by using LAMMPS package[46]. The Stillinger−Weber (SW)[47] potential is used to describe the covalent bonding interaction in MoS$_2$. The interlayer interaction is described by Lennard-Jones (LJ) potential

$$V(r_{ij}) = 4\varepsilon[(\sigma_{ij}/r_{ij})^{12} - (\sigma_{ij}/r_{ij})^{6}] \tag{4}$$

where the parameters are taken from Ref.[48]. The time step is set as 0.5 fs. As shown in Fig. 3(a), the fixed and periodic boundary conditions are adopted along the length and width direction, respectively. To establish a temperature gradient, two Langevin thermostats[49] with different temperature are applied to the two ends of the simulation system. The thermal conductivity $\kappa$ is calculated based on Fourier's Law,

$$\kappa = -\frac{J}{\nabla T} \tag{5}$$

where $\nabla T$ and $J$ are, respectively, the temperature gradient and the heat flux. (More simulation details can be found in the Supplementary Information Part 4.)

The defect concentration is defined as $N_R/N_P$, where $N_R$ and $N_P$ are the number of removed atoms and the total number of atoms in pristine MoS$_2$, respectively. The defects are introduced by randomly removing only sulfur atom in MoS$_2$ considering the mild plasma used in experiment, the relatively light sulfur atoms are much easier to take out. The defect concentration varies from 0.5% to 5%. The spectrum distribution of phonon transmission $T(\omega)$ is calculated as[42]

$$T(\omega) = \frac{q(\omega)}{k_B \Delta T} \tag{6}$$

where $k_B$ is the Boltzmann parameter and $\Delta T$ is the temperature difference between the two Langevin thermostats. $q(\omega)$ is the frequency dependent heat flux across the imaginary cross-section (red dashed line in Fig. 3(a)), which can be calculated as[41, 42]

$$q(\omega) = \frac{2}{t_s} Re \sum_{\alpha \in L} \sum_{\beta \in R} \left\langle \hat{F}_{\alpha\beta}(\omega) \hat{v}_\alpha(\omega)^* \right\rangle \tag{7}$$

where $t_s$ is the simulation time, "L" and "R" respectively denotes the left and right segment, which located at different sides of the imaginary cross-section. $\mathbf{F}_{\alpha\beta}$ is the total force exerted by R segment. (More details are described in Supplementary Information.)

*Minimum thermal conductivity calculation:* We follow the method in Ref.[43] which incorporates both the phonon focusing effect and a first Brillouin zone truncation effect to calculate the in-plane minimum thermal conductivity of amorphous MoS$_2$. The equation is from the Supplementary Information of Ref. [43],

$$\kappa_{\min-ab,Layered} = \sum_{pol} \frac{1}{6\pi v_c} \frac{k_B^3}{\hbar^2} \left[ \begin{array}{c} \int_0^{x_{D,c}} \frac{T^2 x^3 e^x}{(e^x-1)^2} dx + \frac{3}{2}\theta_{D,c} \int_{x_{D,c}}^{x_{D,ab}} \frac{Tx^2 e^x}{(e^x-1)^2} \left( \frac{\theta_{D,ab}^2 - (Tx)^2}{\theta_{D,ab}^2 - \theta_{D,c}^2} \right)^{\frac{1}{2}} dx \\ -\frac{1}{2}\frac{\theta_{D,c}^3}{T} \int_{x_{D,c}}^{x_{D,ab}} \frac{e^x}{(e^x-1)^2} \left( \frac{\theta_{D,ab}^2 - (Tx)^2}{\theta_{D,ab}^2 - \theta_{D,c}^2} \right) dx \end{array} \right] \quad (8)$$

where $x = \hbar\omega/k_B T$. All parameters of MoS$_2$ are taken from the Table 1 in the Supplementary Information of Ref.[50]

**Acknowledgements**

This work is supported by the National Key R&D Program of China (No. 2017YFB0406000), by the National Natural Science Foundation of China (No. 11674245 & No. 11334007 & No. 11774278), by Shanghai Committee of Science and Technology in China (No. 17142202100 & No. 17ZR1447900 & No. 17ZR1448000), and by the Fundamental Research Funds for Central Universities (No. 2012jdgz04)

**Competing financial interest:** The authors declare no competing financial interest.

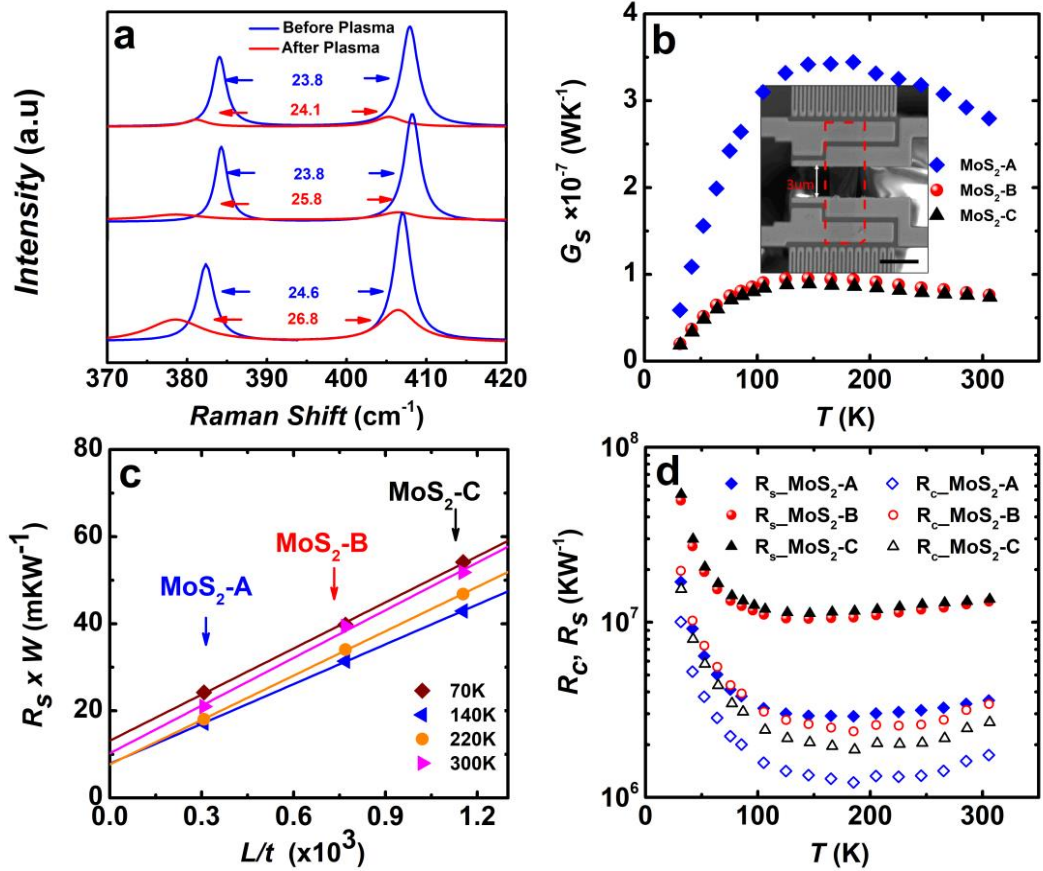

**Figure 1. Sample characterization and thermal conduction measurements.** (a) Raman spectrum for the three samples before (blue curves) and after (red curves) oxygen plasma treatment. The frequency difference (before oxygen plasma treatment) between two Raman bands of $MoS_2$ ($E^1_{2g}$ and $A^1_g$) is 24.6 cm$^{-1}$ and 23.8 cm$^{-1}$ for 5L ($MoS_2$-A) and 4L ($MoS_2$-B and $MoS_2$-C) $MoS_2$, respectively. After Oxygen plasma treatment (red curves), both $E^1_{2g}$ and $A^1_g$ modes are severely depressed with an obvious red shift.[16] (b) Measured thermal conductance of three samples. Inset: SEM image of one typical sample ($MoS_2$-C), the scale bar is 4 $\mu m$. (c) The $R_sW$ versus the $L/t$ for the three samples at different temperatures. The solid lines are linear fitting to the measured data. (d) The thermal contact resistance $R_c$ (unfilled symbols), derived from the intercept of the linear fitting at $L=0$ in (d), and the measured thermal resistance $R_s$ (filled symbols).

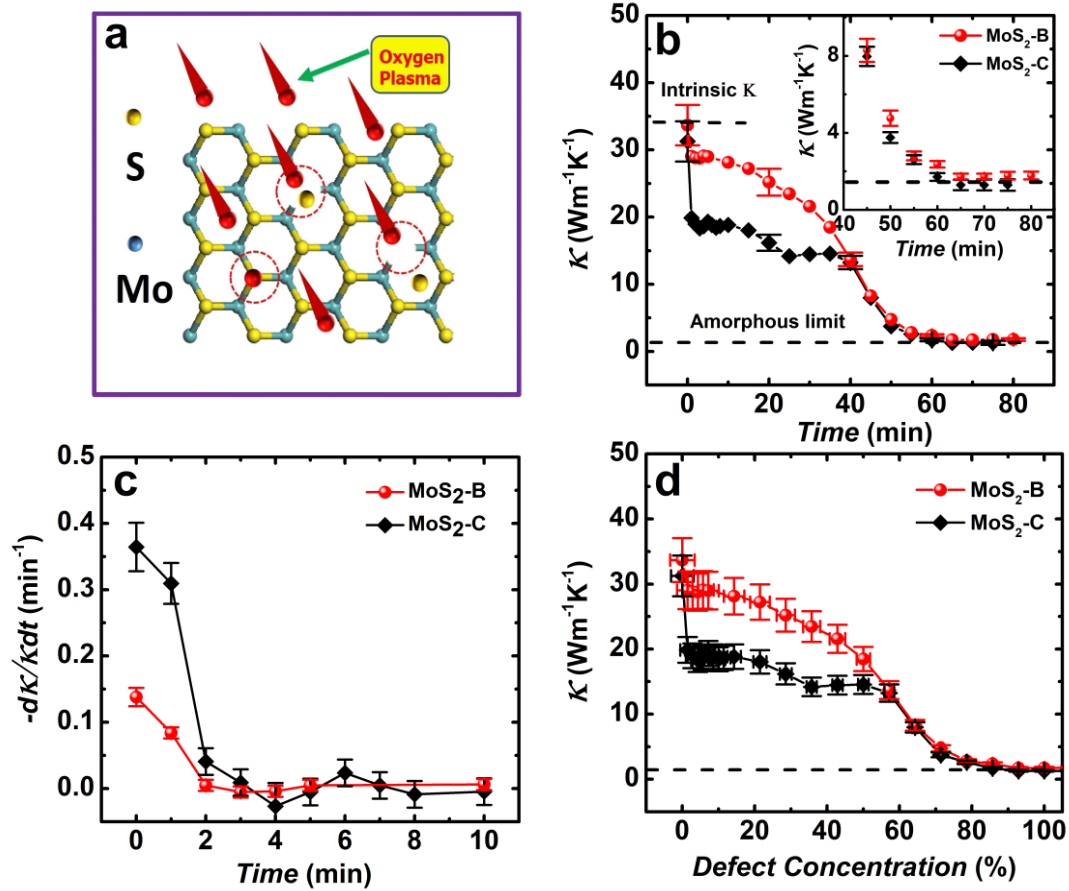

**Figure 2. Tailoring thermal conductivity via mild oxygen plasma.** (a) Schematic of mild oxygen plasma treatment. (b) The measured thermal conductivity versus controlled plasma exposure time. Inset: zoom in on the time range of 45~80 min. (c) The normalized thermal conductivity reduction rate versus exposure time for the first ten minutes. (d) The measured thermal conductivity versus normalized defect concentration (see SI part 3 for details). All the solid curves are plotted to guide the eyes.

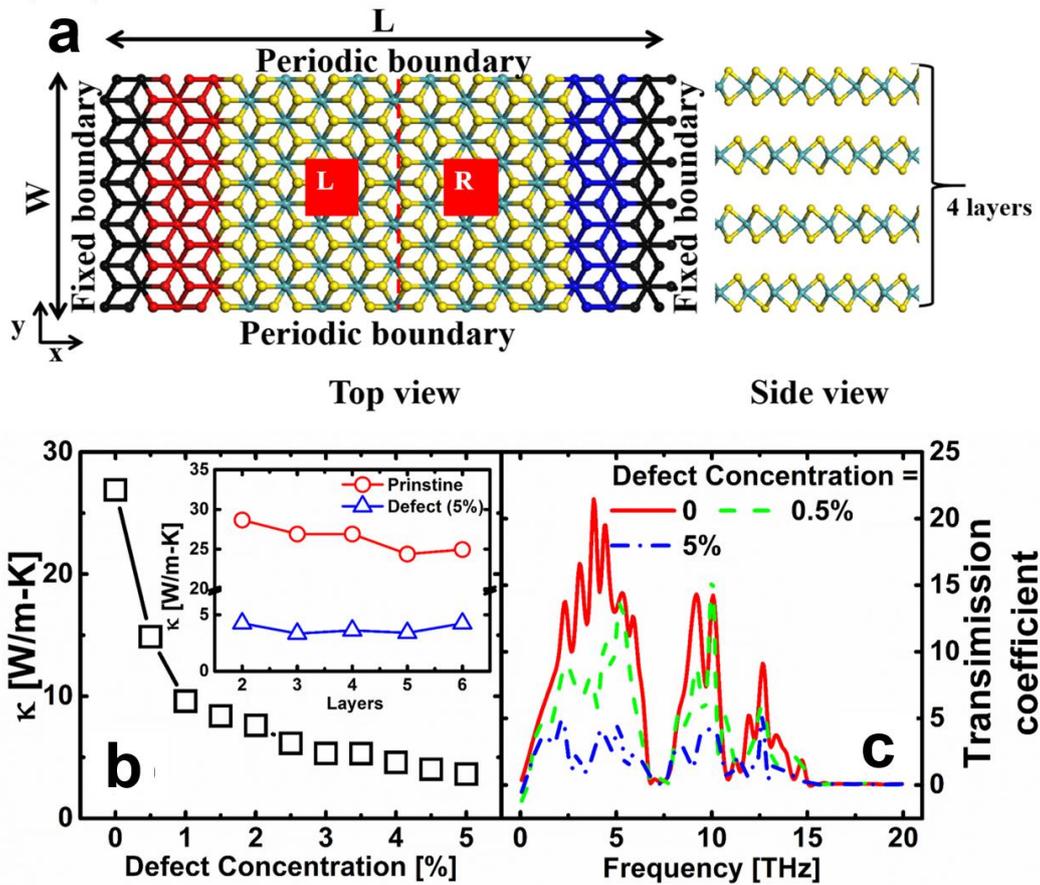

**Figure 3. MD simulations for MoS$_2$ and defected MoS$_2$.** (a) Simulation setup, top view and side view of the four layers MoS$_2$. Fixed boundary conditions are used along the length (*L*) direction, while periodic boundary conditions are used along the width (*W*) direction. In the MD simulations, the size of the simulation domain is fixed as *L*= 50 nm and *W*=5 nm. The red dashed line denotes the imaginary plane which divides the system into "L" and "R" two segments. (b) Thermal conductivity of MoS$_2$ versus the defect concentration at room temperature. The thermal conductivity of MoS$_2$ decrease with the defect concentration increase. Inset: effect of defects concentration and number of atomic layers on the thermal conductivity of MoS$_2$ predicted by the MD simulations. The red circles denote pristine MoS$_2$ and the blue triangles denote defective (5%) MoS$_2$. (c) The transmission coefficient of pristine MoS$_2$ and defected MoS$_2$ with different defect concentration.

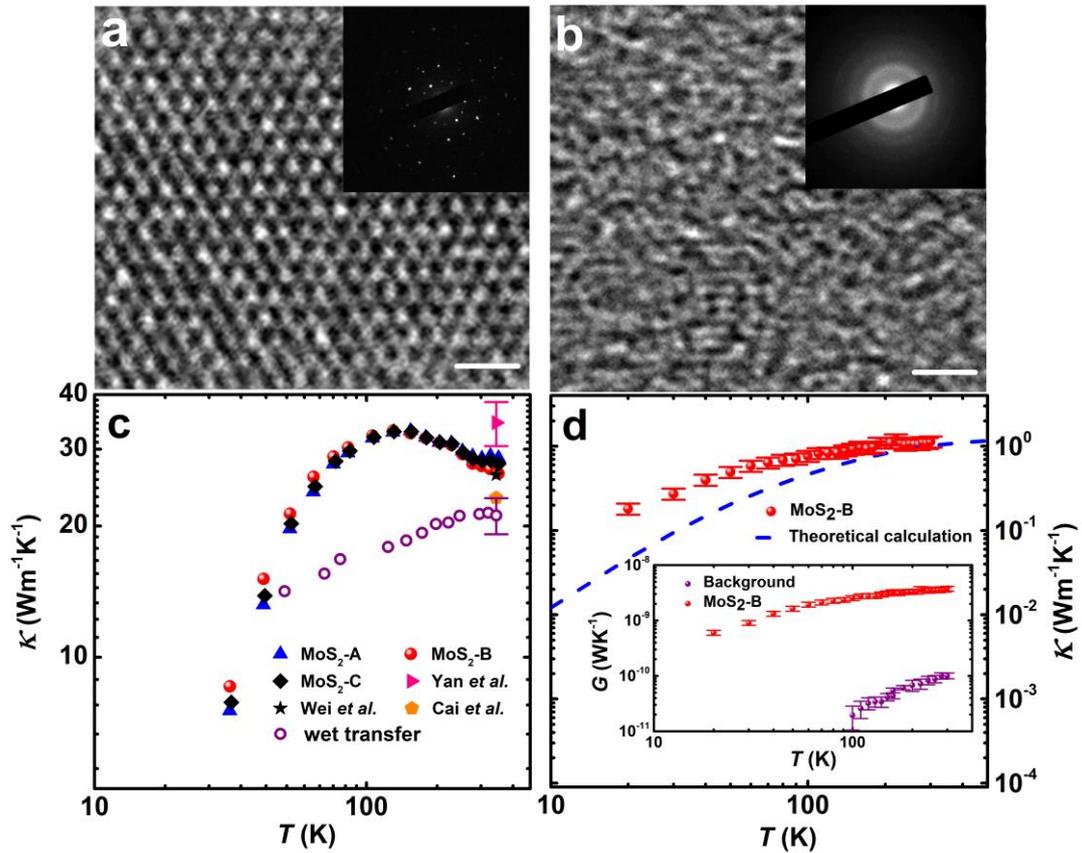

**Figure 4. Characterization of samples in crystalline and amorphous phase with TEM and thermal measurements.** (a) HRTEM image and diffraction pattern of the intrinsic sample indicative of crystalline phase. (b) HRTEM image and diffraction pattern of the tailored sample indicative of amorphous phase. The scale bar is 1 *nm* (c) The derived thermal conductivity (before plasma treatment) of the samples as a function of temperature. Shown for comparison are the measured room-temperature values of a monolayer $MoS_2$ sample (pink triangle) reported by Yan *et al.*[19] and the thermal conductivity calculated by Wei *et al.*[31] (black star) and Cai *et al.*[32] (orange pentagon), respectively. The measured thermal conductivity of few layer $MoS_2$ prepared by wet transfer method (purple open circles) is also plotted to compare with the one prepared by dry transfer method. The measured thermal conductivity is much smaller than that in three other samples with no obviously peak when temperature is below *T*=300K. This is due to the additional scatterings from organic on surfaces, indicate that samples prepared by dry transfer method have a much superior quality. (d) The measured thermal conductivity in the range of 20 K to 300 K after the oxygen plasma process. The blue dashed line denotes the theoretical low limit of amorphous thermal conductivity in $MoS_2$. Inset for (d): the measured thermal conductance of $MoS_2$-B and background heat conductance of the blank suspended device.

**Table 1. Parameters of samples used for thermal measurement**

| Sample names | Number of layers | Width ($\mu m$) | Length ($\mu m$) |
|---|---|---|---|
| $MoS_2$-A | 5 | 5.87 | 1 |
| $MoS_2$-B | 4 | 3 | 2 |
| $MoS_2$-C | 4 | 3.82 | 3 |

## Notes and References